\documentclass[twocolumn]{aastex701}

\newcommand{\etal}{\mbox{\rm{et al.}~~}}
\newcommand{\kms}{\mbox{km\,s$^{-1}$}}
\usepackage{color}
\usepackage{changepage}
\usepackage{float}
 \usepackage{tikz}
\begin{document}

\title%{Accelerating massive galaxy formation}%Self interacting or primordial dark matter ?}
{Accelerating massive galaxy formation with primordial black hole seed nuclei}
%% AUTHOR/INSTITUTIONS FOR AASTEX6.1:
%% AUTHOR/INSTITUTIONS FOR AASTEX6.1:
\author{Jeremy Mould}%

\email[show]{jmould@swin.edu.au}  
%\altaffilmark{1,2}
\affiliation{Swinburne University}%$^1$}
\affiliation{ARC Centre of Excellence for Dark Matter Particle Physics}%$^2$}
%% AUTHOR/INSTITUTIONS FOR AASTEX6.1:

%% Use the \collaboration command to identify collaborations. This command
%% takes an optional argument that is either a number or the word "all"
%% which tells the compiler how many of the authors above the command to
%% show. For example "\collaboration[all]{(DELVE Collaboration)}" wil include
%% all the authors above this command.
%%
%% Mark off the abstract in the ``abstract'' environment. 
\begin{abstract}
If massive primordial black holes (PBHs) exist and constitute a fraction of the dark matter, they can dramatically catalyze galaxy formation. By acting as pre-existing, high-density seeds, they can shorten the galaxy assembly time to as little as 100 Myr
 for up to 10$^8$ M$_\odot$ PBH seeds, %(Hoshi \& Yamada 2025)
allowing for the rapid formation of host halos. %The subsequent hierarchical merging of these seeded halos %, i%modulated by baryonic feedback, 
        %naturally gives rise to a galaxy luminosity function that resembles the observed Schechter function.
Furthermore, low surface brightness or diffuse galaxies may represent a natural outcome of this process, perhaps as the residue of halos seeded by smaller PBHs that failed to accrete a major baryonic component. %If 
%Massive PBHs  can catalyse galaxy formation.

\end{abstract}

%% Keywords should appear after the \end{abstract} command. 
%% The AAS Journals now uses Unified Astronomy Thesaurus (UAT) concepts:
%% https://astrothesaurus.org
%% You will be asked to selected these concepts during the submission process
%% but this old "keyword" functionality is maintained in case authors want
%% to include these concepts in their preprints.
%%
%% You can use the \uat command to link your UAT concepts back its source.
%\keywords{\uat{Galaxies}{573} --- \uat{Cosmology}{343} --- \uat{High Energy astrophysics}{739} --- \uat{Interstellar medium}{847} --- \uat{Stellar astronomy}{1583} --- \uat{Solar physics}{1476}}
%% From the front matter, we move on to the body of the paper.
%% Sections are demarcated by \section and \subsection, respectively.
%% Observe the use of the LaTeX \label
%% command after the \subsection to give a symbolic KEY to the
%% subsection for cross-referencing in a \ref command.
%% You can use LaTeX's \ref and \label commands to keep track of
%% cross-references to sections, equations, tables, and figures.
%% That way, if you change the order of any elements, LaTeX will
%% automatically renumber them.
\section{Introduction}
One of the most striking early outcomes from the James Webb Space Telescope (JWST) has been the discovery of an unexpected abundance of UV-bright and massive galaxies at very high redshifts (z $>$ 8), seemingly in tension with standard galaxy formation models (Ma \etal 2025, de Graaff \etal 2025, Xiao \etal 2024). 
P\'erez-Gonz\'alez \etal (2025)  find the need for an enhanced UV-photon production at z $\approx$ 20 in  $\sim$10$^9$ M$_\odot$ dark matter halos, provided by an increase in the star formation efficiency at early times and/or by intense compact starbursts with enhanced emissivity linked to strong burstiness, low or primordial gas metallicities, and/or a top-heavy initial mass function.
This is perceived as a significant challenge for the 
standard $\Lambda$CDM cosmology (Merlin \etal 2024, Boylan-Kolchin 2025
), where structure forms hierarchically and requires substantial time to build massive objects. While not a fatal flaw -- as modifications to the assumed star formation efficiency or stellar initial mass function at early times may resolve the tension (Yung, Somerville \& Iyer 2025; Ziegler \etal 2025) -- it motivates exploring alternative scenarios. If a significant fraction of dark matter (DM) exists as primordial black holes (PBHs), this provides a potent mechanism to accelerate early structure formation, 
as Inman \& Ali-Ha\"imoud (2019), Capelluti (2023) and Liu \& Bromm (2022)  have also discussed.
Zhang, Liu \& Bromm (2025) find that PBH host halos, through gravitational influence, significantly impact the structure formation process, compared to the CDM case, by attracting and engulfing nearby newly-formed minihalos.
Delos \etal (2024)  identify numerous dynamical effects due to the collisional nature of PBH dark matter.
\section{Galaxy formation with PBH}
PBHs meet the basic criteria for a good dark matter candidate. They are effectively collisionless, non-baryonic, as they do not carry a baryon number, and formed before Big Bang Nucleosynthesis (t 
$<$ 1 s), and, for masses above $\sim$10$^{15}$ g, are stable on timescales far exceeding the age of the Universe. They are not entirely dark, as they emit Hawking radiation with a luminosity 
L $\approx$ --c$^2$dM/dt, where dM/dt is the  mass loss rate,
%L = ħc⁶/(15360πG²
 but this is only significant for low-mass PBHs and is negligible for massive objects. The concept that PBHs could form from the collapse of large primordial density fluctuations ($\delta\rho/\rho~\gtrsim$ 1) was first explored by Zel'dovich and Novikov (1967) and Hawking (1971). Recent comprehensive reviews are provided by Green (2024) and Carr \& K\"uhnel (2022).

Although LIGO is so far equivocal about
massive PBH (M $>~$ 10$^4$ M$_\odot$), they are considered here as seeds for galaxy formation. To achieve the local DM %dark matter density 
of $\rho_{DM}~\approx$ 0.4 GeV/cm$^3$ 
($\sim$0.01 M$_\odot$/pc$^3$), the required number density of such objects would be very low. For instance, if the DM were composed entirely of 10$^6$ M$_\odot$ PBHs, their local number density would be n $\sim$10$^{-7}$ /pc$^3$, implying the nearest such object would likely be just kiloparsecs away. %Their large velocity dispersion (~200 km/s) and minuscule number density make their non-detection via direct dynamical influence on solar system bodies entirely unsurprising (Mould 2025a).
%One of the early outcomes of JWST has been pressure on galaxy formation models to develop galaxies quickly (Ma \etal 2025,
%de Graaff \etal 2025, Xiao \etal 2024). This is seen as a problem for
%the standard $\Lambda$CDM cosmology (Merlin \etal 2024),
%although not a fatal one (Yung, Somerville \& Iyer 2025).

The mass range of PBHs is potentially very large, ranging from a Planck mass to 10$^7$ M$_\odot$ (
Mould \& Batten 2025). Here I consider two masses: subsolar mass
PBHs, which may make up a significant fraction of the DM (Tran \etal 2024),
and supermassive PBH, a natural nucleus for a galaxy. Evolutionary tracks from the radiation dominated era to the current epoch are
presented by Mould (2025b). Observational evidence for their existence, however, is ambivalent (K\"uhnel 2025).
While lensing surveys like OGLE have probed for PBHs in the asteroid-to-stellar mass range, with candidate events reported by Niikura \etal (2019) 
 and Sugiyama, Takada, Yasuda \&  Tominaga (2026),
PBH with a mass in the range 10$^{-11}$ to 10$^{-5}$
M$_\odot$ %by Key \etal (2024)
are candidates for a significant fraction of the DM.
To reach the local density of DM 0.01 M$_\odot$ pc$^{-3}$ would require a number density up to 10$^{13}$ pc$^{-3}$ i.e. of order 1 within Neptune's orbit at any time, moving with a speed of at least 100 \kms, and yet none have been identified (Mould 2025a) by direct detection.

In this paper a focus is on the LIGO mass to supermassive end of the scale.
There is extensive literature on supermassive PBHs, starting with Bicknell \& Henriksen (1979), and reviewed
by Mould \& Batten (2025). Imai \& Mathews (2025) and Mould \& Hurley (2025) have presented simulations in which intermediate mass PBH
exceeding 1000 M$_\odot$ form by accretion preceding galaxy formation.
Boylan-Kolchin (2025) has also found that dark matter 
%and standard cosmological evolution may therefore 
may be crucial for explaining the surprisingly high levels of star formation in the early Universe revealed by JWST. 

What are the limits on the numbers of supermassive black holes (SMBH) at the
time of galaxy formation ? From quasars at z $\sim$ 6 we have n $\sim$
10$^{-6}$ Mpc$^{-3}$ (Shen \etal 2020). This is a fraction 1.5 $\times$ 10$^{-20}$ of the 
DM at that redshift. From galaxies of the size of the Milky Way
the Schechter luminosity function gives us 4.2 $\times$ 10$^{-3}$ Mpc$^{-3}$
at the current epoch. The mass of Sgr A implies a fraction   6 $\times$ 10$^{ -7}$ of the DM. From globular clusters in the Local Group it is possible to put
an upper limit %630 $\times$ 10$^4$ / 2 $\times$ 10$^{12}$ = 
of 3 $\times$ 10$^{-6}$ for the fraction of the DM in 10$^4$ 
M$_\odot$ intermediate mass black holes (IMBH), supposing every cluster has one of 
these as its nucleus
by multiplying
their number by 10$^4$ and dividing by the mass of the Local Group. 
 None of these estimates violate calculated limits on SMBH
and IMBH PBHs found by Poulin \etal (2017) for accreting PBHs: 10$^{-4}$
in the accretion disk sound speed, c$_s$ limited case, and 10$^{-5.5}$
in the $\surd c_s$ limited case for M$_{PBH}$ = 10$^3$ M$_\odot$. The hypothesis in that paper is that the
non-ionizing Hawking radiation of such PBHs is supplemented by keV radiation from
an accretion disk.%}

In $\S$2 some DM-only simulations are carried out,  using
an n-body calculation of their mutual gravitational attraction, and in $\S$3  galaxy formation time is considered. 
%A galaxy mass function follows in $\S$4, equivalent to a luminosity function for an old stellar population. 
Also in $\S3$ diffuse galaxies are 
hypothesized to be the residue of this process.
%A number of 100000 particle n-body runs were made with the dark matter only
%code of Mould \& Hurley (2025). %whose details are in Table 1 and illustrated in Figure 3. 

\begin{figure}	[H]
\includegraphics[width=.6\textwidth,angle=- 0]{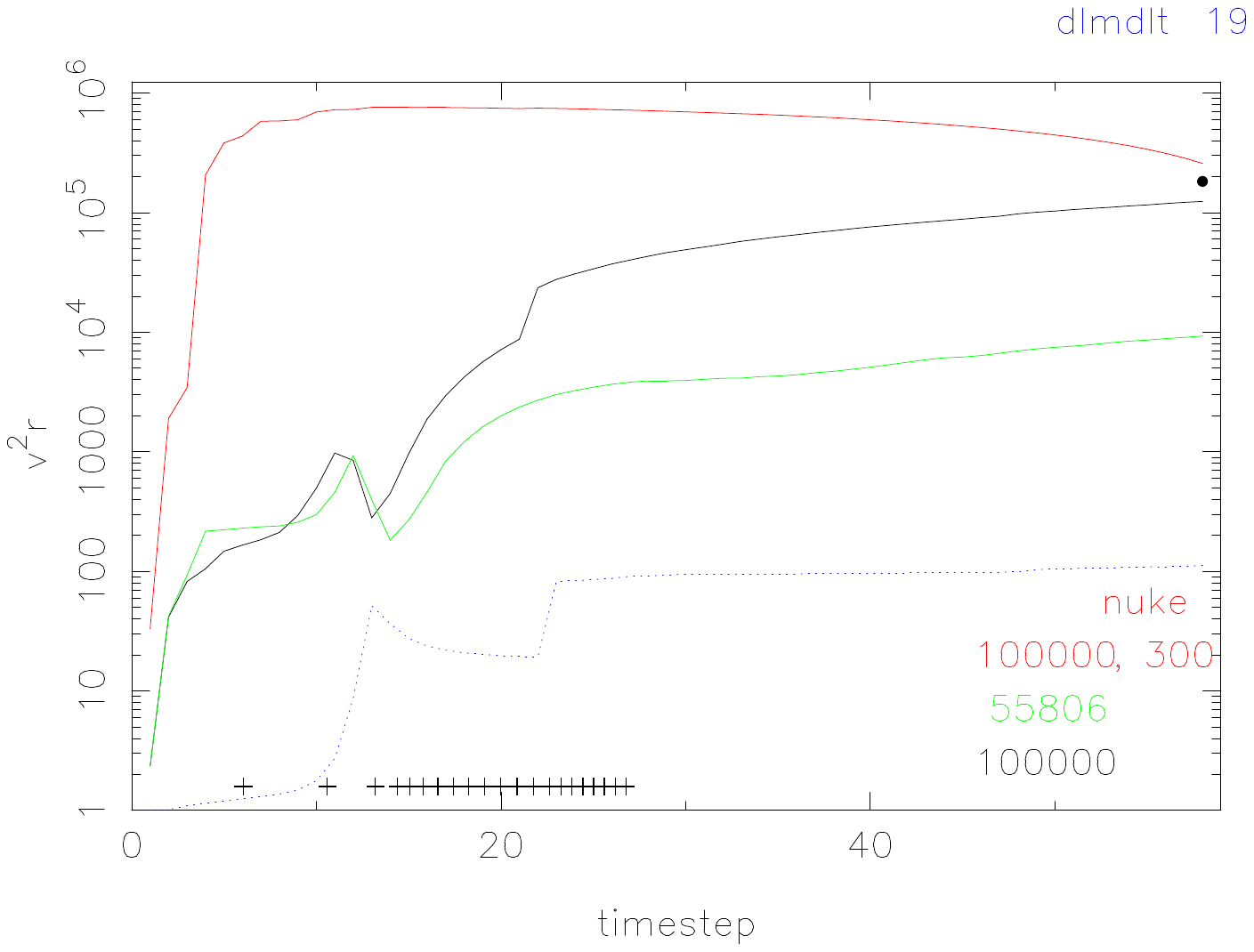}
\caption{
	The green line is the evolution in $r_e\sigma^2$ of 100,000 PBH %dark matter 
	particles that lose 1\% of their mass per time step , to be compared with mass conserving particles (black line). Both axes are dimensionless.  The x-axis is time %the radius vector from the nucleus
	; the y-axis, $r_e \sigma^2$, velocity dispersion (v or $\sigma$) squared multiplied by the effective radius. The
	effective radius contains half the mass. The dimensionless mass loss rate dlogM/ dlogt is shown by the dotted blue line. It averages 19 (noted at the top right), which is higher than PBH emit for most of their lives. The red line shows that a nucleus with a mass as low as 300 particles energises the early time steps of a simulation. Uneven time steps were used , as shown by the distribution of the plus signs. The bumps in the curves are not numerical problems; they are an artefact of the sharp outer boundary of the initial uniform distribution of particles and correspond to the crossing time. The number of particles in the simulation is noted at the bottom right. 
The point to the right is a test of sensitivity to the initial distribution. Instead of uniform, a power law in radius was adopted. The point is placed at the end that run.}
\end{figure}

Tracks of PBH evolution by
Mould (2025b) show that PBH between 10$^{-20}$ and 10$^{-18}$ M$_\odot$ %m20 and m18 
evaporate between the z$_{eq}$
and z = 100. For present purpose we simulated their contribution (and that of the
next several decades in mass) to the evolution of structure with the DM-only code of Mould \& Hurley (2025) and 100,000 PBH %dark matter 
particles. This code, like all n-body codes, calculates
the positions of all the particles by determining their mutual accelerations,
and thus their velocities.
The initial conditions were a spatially uniform random distribution of particles in a sphere with a nucleus of mass M, the largest in the distribution and zero velocity.  PBH particles ranged in mass between m1 and m2, nominally 0.1 to 1 M$\odot$, but the simulations are scale free, and their ratios are the relevant quantities. The mass distribution
was n $\sim$ m$^{-1}$ placing equal mass in each equal logarithmic interval.
%The simulations are scale free in radius. %, but accretion is density dependent (dm/dt =
Mass loss was simulated in a scale free way by losing 1\% of the particles per timestep.  Similar results are obtained applying dm/dt to the masses each timestep. A dimensionless measure of mass loss is dlogM/dlogt. This is shown in Figure 1. PBH lose mass at a rate that is initially zero and rises to r = dlogM/dlogt = 7/3 when half their mass is gone. After that their evaporation is rapid, as r $\sim$ M$^{-3}$.

The resulting structure of the DM %dark matter 
halos is shown in Figure 2. The early formation and subsequent mass loss create halos that are more centrally concentrated for a given mass compared to standard $\Lambda$CDM. Halos with these deeper potential wells possess higher escape velocities, making them better able to retain baryonic gas in the event of energetic feedback from supernova explosions or AGN. 
This has significant implications for chemical evolution, as retaining gas allows for multiple generations of star formation and progressive metal enrichment. It is unlikely that 100\% of the Universe's DM (f$_{PBH}$ = 1) is composed of PBHs in this narrow mass range, so the order-of-magnitude effect we observe on the velocity dispersion (r$_e\sigma^2$) should be considered an upper limit, with the real effect being scaled by the true f$_{PBH}$.
%The radius $r_e$ is the half mass radius, and $\sigma$ is the velocity dispersion.

%We find that  ($r_e\sigma^2)_1/(r_e\sigma^2)_0$ = 0.1 for  1\% mass loss, where the zero subscript signifies no mass loss. The radius r$_e$ is the half mass radius. The kinetic energy per unit mass (and relative depths of the halo potentials) $\sigma^2_1/\sigma^2_0$ = 0.51. The simulation was stopped when 45\% of the PBH had evaporated. The redshift then was 20\% smaller than when the simulation began. %z$_{eq}$ = 3400 in the Planck (2020)  cosmology. 
%The value for a halo with no mass loss and half the  number of initial particles was ($r_e\sigma^2)_2/(r_e\sigma^2)_0$ = 0.125, where the subscripts 0 and 2 signify no mass loss. 
%M~v_circ^n ; v_circ=0.83*f/0.5 or function(f) TBD with sims
\begin{figure}[H]	
%\vspace{-3 in}	
%\includegraphics[height=15cm,width=1.35\textwidth]{colorbar2.pdf}
 %\vspace{-1in}
\hspace{-.5in}
\includegraphics[width=.7\textwidth]{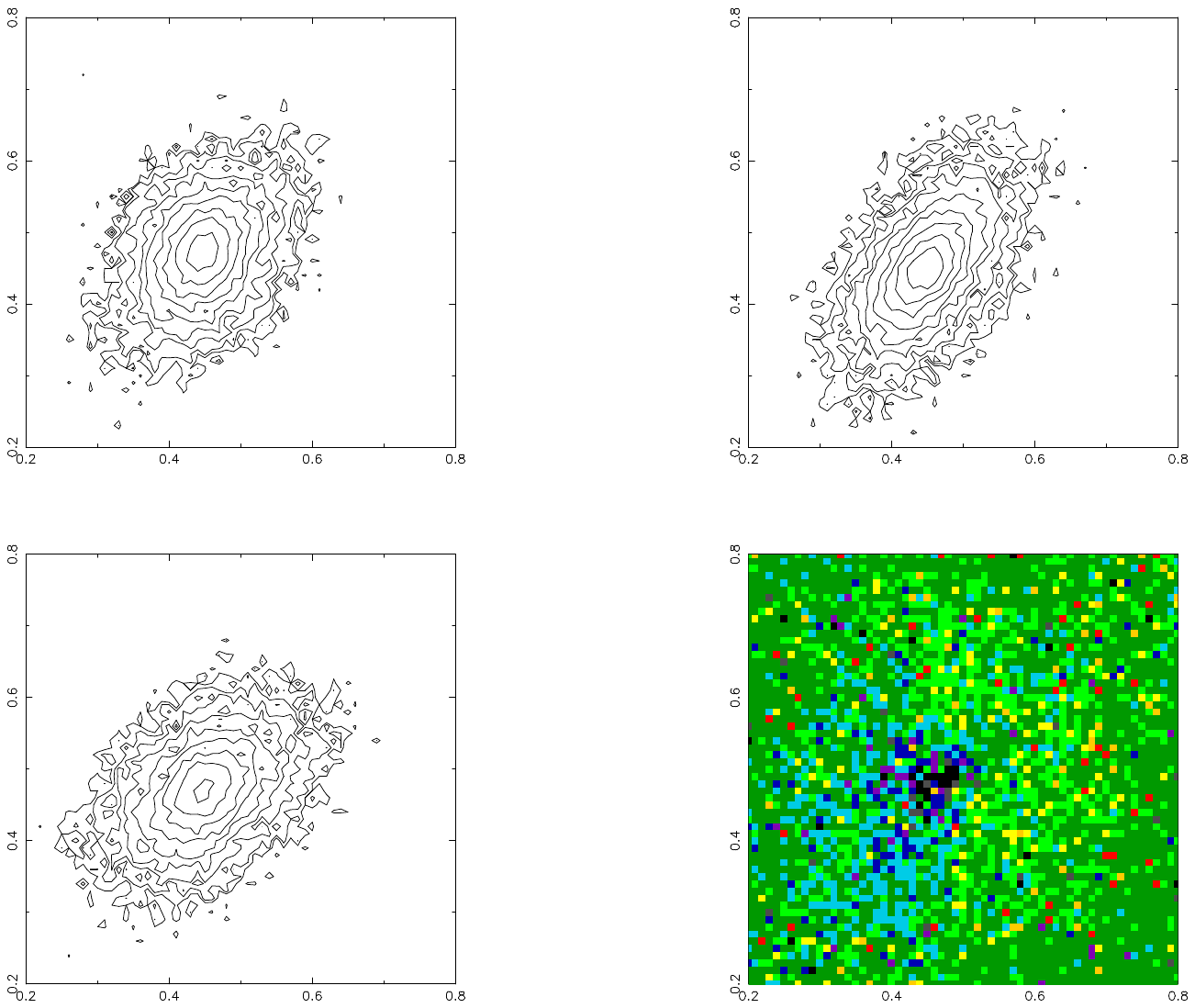}
 \vspace{-1cm}
\includegraphics[width=.6\textwidth, clip, trim=9cm  5cm 1cm  8cm]{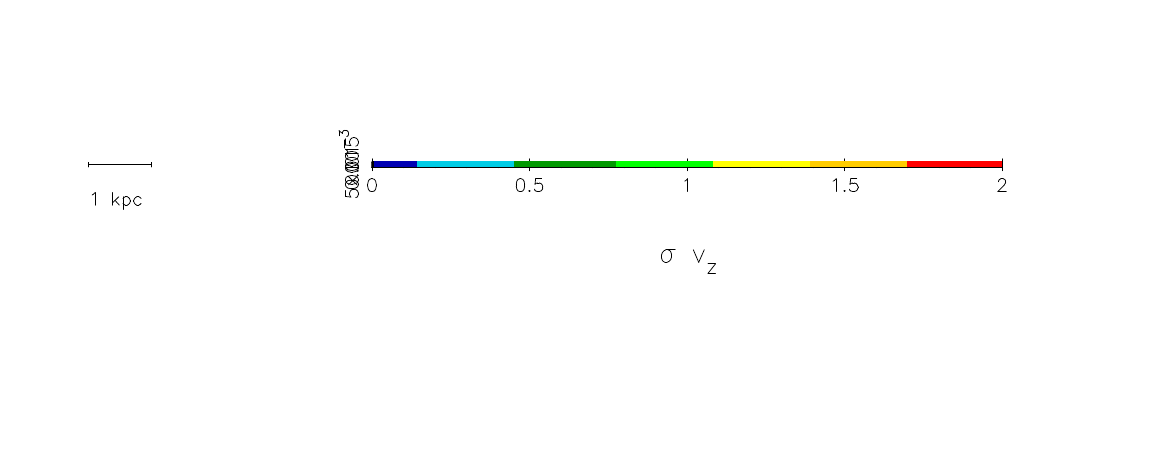}
 \vspace{-1.5cm}
	\caption{The x-y, y-z, and x-z projection density contours of a 1\% mass loss per timestep simulation.  The contours are the loci
	of equal projected mass density,  separated by 0.4 dex. Early formation creates halos
	that are more centrally concentrated.
	%The scale on the axes is pixels. 
	Elliptical contours are common. The color figure is velocity in the z direction from the x-y projection.
	Blue is approaching; the red points are redshifted.  One unit in the color bar
	above is the v$_z$ velocity dispersion.}
\end{figure}
\subsection{Supermassive PBH}
%At the other end of the PBH mass range are supermassive black holes (SMBH). The presence of SMBH at redshifts accessible to JWST make a primordial origin
%worth entertaining (but see Kusenko 2025 for an alternative). Mould \& Batten (2025) find that a PBH origin yields an excellent fit to the QSO
%luminosity function.
%\pagebreak
%{\color{red} The next section shows how  v$_{circ}$ affects the LF.}
At the other end of the PBH mass spectrum are supermassive black holes (SMBHs). The presence of $\sim$10$^8$ M$_\odot$ quasars at redshifts z $>$ 6, as confirmed by JWST (Hoshi \& Yamada 2025), presents a significant timing problem for models that grow black holes from stellar-mass seeds. A primordial origin for these seeds is therefore worth entertaining (but see Kusenko (2025) for alternatives like direct collapse black holes). Mould \& Batten (2025) find that seeding galaxies with massive PBHs yields an excellent fit to the observed QSO luminosity function. Simulations with massive PBH seeds are shown in Figure 3. 
While halos seeded by intermediate-mass PBHs can show cored profiles, those seeded by 10$^6$ and 10$^7$ M$_\odot$ PBHs develop steep central density cusps
\footnote{As Jiang \etal (2025) have found, the introduction of self-interacting dark matter (SIDM) can subsequently erase these cusps, as frequent particle scattering in the dense nuclear region thermalizes the inner halo and flattens the density profile.
} due to the gravitational focusing of ambient DM %dark matter 
on to the central seed. 

\subsection{The PBH initial mass function}
%An IMF of n $\sim$ m$^{-1}$ produces galaxy-like radial profiles (Figure 4).

It is physically unlikely that PBHs would form as a monochromatic (single-mass) population. A more realistic scenario involves an initial mass function (IMF), or mass spectrum, determined by the shape of the primordial power spectrum. Extended mass functions have been considered by Carr \etal (2016) and Mould (2025b). An IMF with a power-law form, such as dN/dM $\propto$ M$^{-1}$, leads to a hierarchy of seed masses. Our simulations show that such a spectrum, after undergoing hierarchical assembly, can produce a population of halos with radial density profiles that broadly resemble the Navarro-Frenk-White (NFW) profiles of observed galaxies (Figure 4).
\begin{figure}%[h]
%\hspace{-2cm}
\includegraphics[width=.6\textwidth]{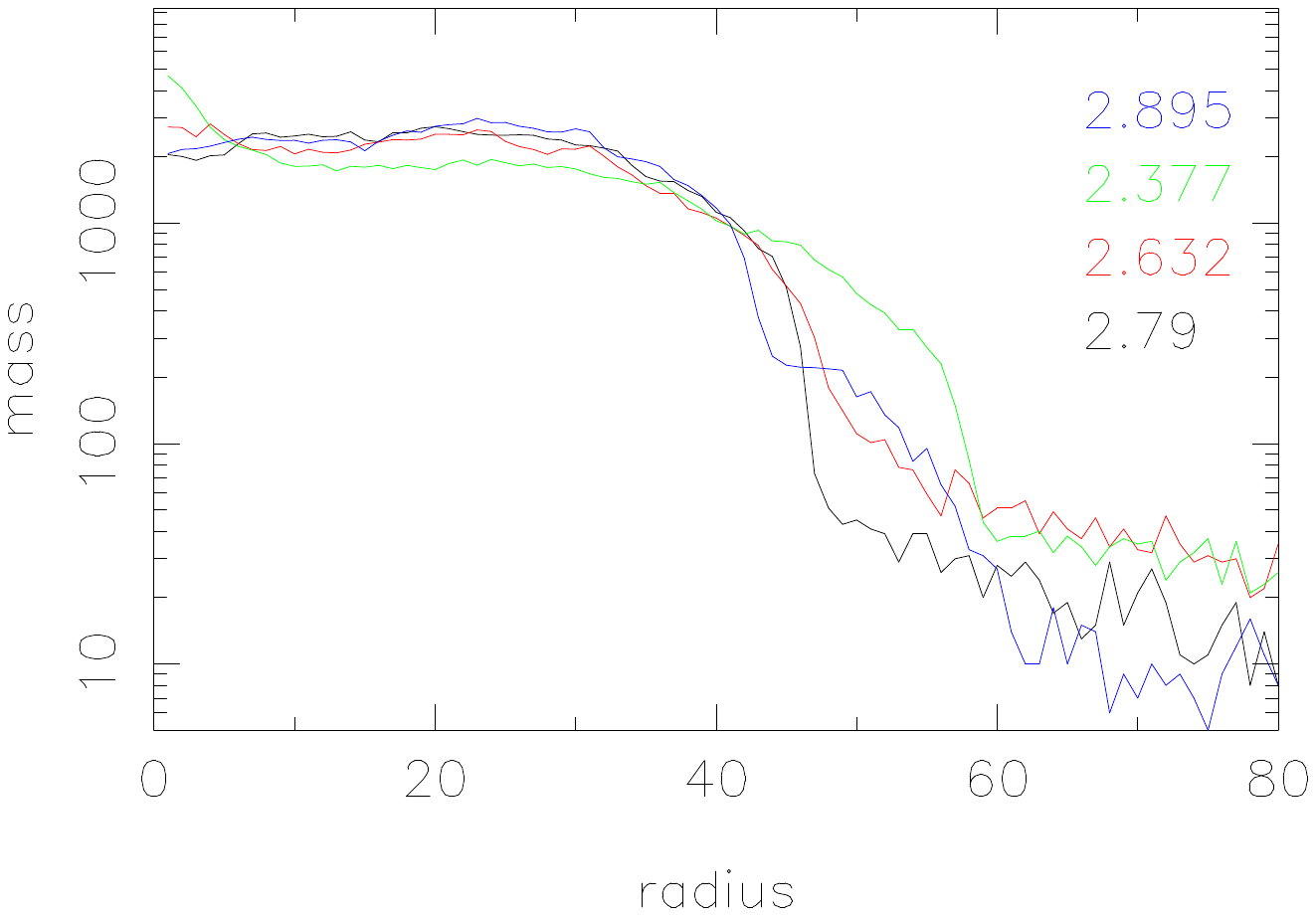}
	\caption{Final radial profiles of simulations with supermassive PBH nuclei.  These are calculated by azimuthally averaging the DM density. The colored numbers on the right are
	effective radii. The x-axis is pixels, the y-axis solar masses per pixel. Nuclei of 10$^5$, 10$^6$ and 10$^7$ M$_\odot$ are shown in black, red and green respectively.   These are runs 53, 61 \& 132 in Table A2. Although these simulations are scale free,
	once these masses are assigned to the PBH nuclei, pixel units
	and timescales follow. Effective radii are nominally in kpc. In blue are shown a 10$^6$ M$_\odot$ case with self-interacting DM. Adding a scattering cross section
	removes the cusp. This is not a property associated with PBHs.}
\end{figure}
\begin{figure}%[h]
\hspace{-1cm}
\includegraphics[width=.6\textwidth]{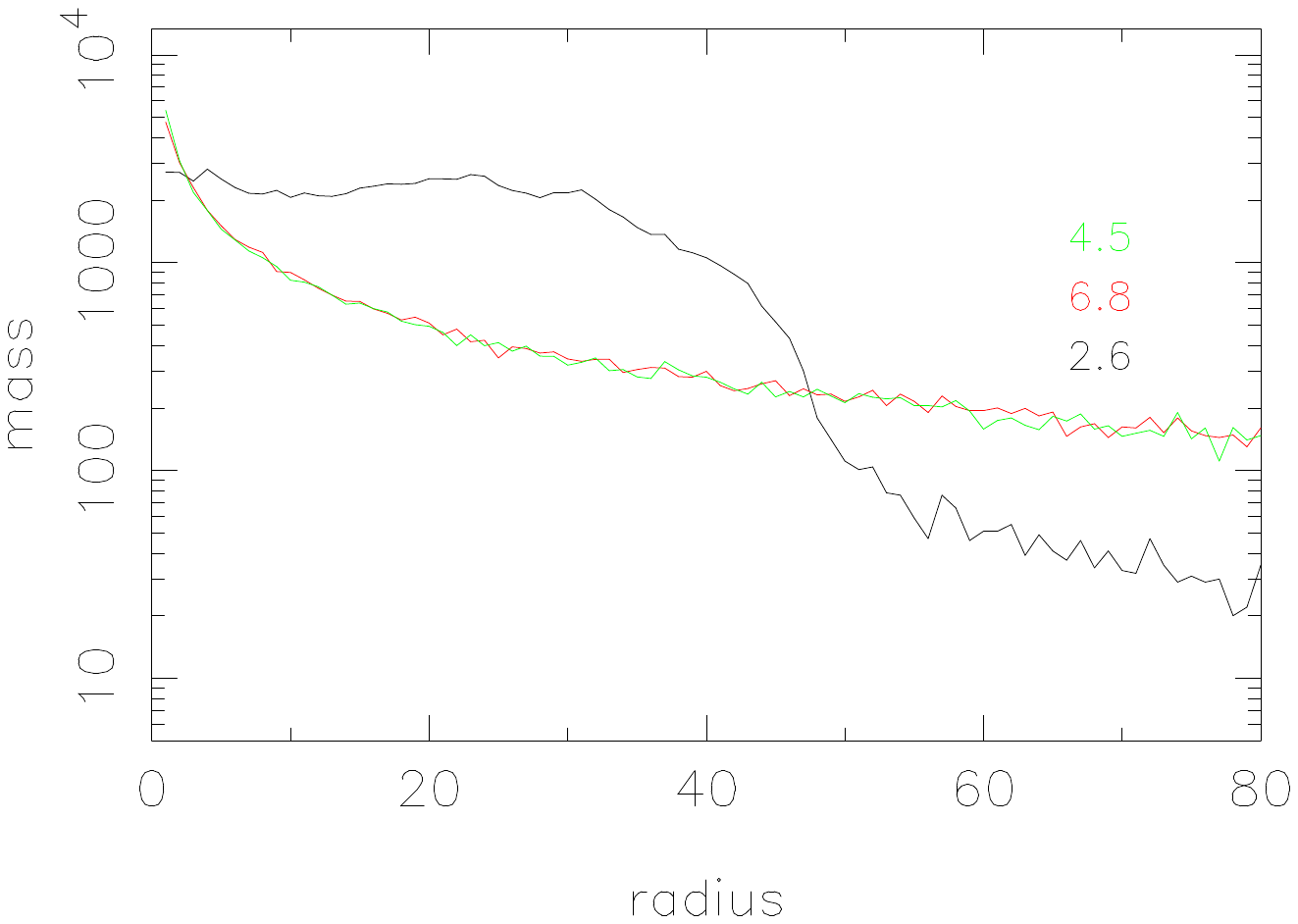}
\caption{Final radial profiles of simulations with supermassive PBH nuclei as in the previous figure. A nucleus of 10$^5$ M$_\odot$ is shown in black  (run 187). Effective radii, nominally in kpc, are also noted. In green and red we show  10$^4$ and 10$^6$ M$_\odot$ (runs 185 and 188)  cases with 	an N $\sim$ M$^{-1}$ IMF.
	 The colored numbers
	are effective radii. The difference between the cored
	profile and the other two is a gap in the mass function
	between the nucleus and the next PBH mass.}
\end{figure}
Cusps or cores as seen in Figure 4 are liable to be affected by numerical artifacts (e.g., softening length, discreteness noise, or two-body relaxation, Binney
\& Knebe 2002). Discreteness noise is visible in our simulations with particle number below 100000, as listed in Table A1. The stability of cusps is not necessarily
an issue in the present context, where we are interested in the early effects of massive nuclei. Also omitted in the limited simulations presented here is the effect of PBH
accretion, which can enlarge nuclei over sufficient time. Instead, we draw on the simulations of Mould \& Hurley (2025) who found that nuclei of 1--10 $\times$ 10$^3$ M$_\odot$
grew by Bondi accretion to masses of 1--2.5 $\times$ 10$^4$ M$_\odot$. There is general agreement in the literature on accretion: Prole \etal (2025) find that at f$_{PBH}$ = 10$^{-3}$ , a number of PBHs were able to embed themselves in dense gas and grow to 10$^4$ - 10$^5$ M$_\odot$ by z = 20. Zhang \etal (2025) find that PBH accretion is self-regulated by feedback, suppressing mass growth unless feedback is weak. PBHs accelerate structure formation by seeding dark matter halos and gravitationally attracting gas, but strong feedback can delay cooling and suppress star formation. %In addition, the presence of baryon–dark matter streaming creates an offset between the PBH location and the peaks induced in gas density, promoting earlier and more efficient star formation compared to standard ΛCDM. 
By z $\sim$ 10, PBH-seeded galaxies form dense star clusters. 
\section{The galaxy formation time}%luminosity function}
In the presence of a central PBH mass M with potential  V $\sim$ GM/R
%At their PBH formation time and temperature of the universe T (for mass M = 3.3 $\times$ 10$^{6}$ (MeV/kT)$^2$; Mould 2025b),
the number density of protogalaxies with PBH nuclei is  one per volume
occupied by a Schwarzschild radius r, $$n = 3/4\pi /r^3 = (3/4\pi) c^6/G^3m^3 \eqno (1),$$ 
where r is the Schwarzschild radius and m the mass. 
%We can then estimate the collision rate per unit volume for these PBHs, which is given by $\Gamma$ = n$_{PBH}^2 <\sigma v>$, where $<\sigma v>$ is the velocity-averaged cross-section\footnote{ For PBHs, the relevant cross-section is not their geometric size but the larger gravitationally focused cross-section, $\sigma \approx  r^2 (1 + v_{esc}^2/v^2$), where v is the relative velocity. We
%omit this correction factor.}
%The
%collision rate per unit volume is %the geometric cross section times the velocity, v,  and number density squared 
%$$N_{coll} = \frac{\pi G^2M^2/c^4 (3/4\pi)^2 c^{12}}{G^6M^6} v = 0.75^2 v c^8/G^4M^4 /\pi \eqno (2),$$
%For 10$^8$ M$_\odot$ PBHs this is 0.75$^2/(1.5 \pi ~\times~ 10^{13})$ v AU$^{-3}s^{-1}$,
%i.e. by the time of recombination the time between collisions, t$_1$ = 3.2 $\times$ 10$^7$ years per AU$^3$ for v = 200 \kms.
%We shall see in the next section that, if the equation (2) rate is
%multiplied by a Hubble time, we obtain the number density of galaxies per unit luminosity.
%{\color{red}
In an $\Lambda$CDM universe the relation between a
bound structure's mass and radius R is (Kaiser
1986; Padmanabhan 1993; Somerville \& Primack 1999)
R~$\propto~M^\gamma$ 
with $\gamma$ known in terms of the power-spectrum of
density fluctuations P (k) at wavenumber k. 
If 
the power spectrum is fitted locally to a power-law of index n$_k$, P (k) 
$\propto~ k^{n_k}$ , the index shifts progressively
from n$_k$ = +1 to n$_k$~=~-3 as one moves to smaller
scales. The indices n$_k$ and $\gamma$ are linked through
the mass-radius relation 
$\gamma$ = (n$_k$ + 5)/6.
%Note that γ is related to the power-index ν relating
%mean density and mass, ρ ∝ M ν , by ν = γ − 3 =
%(nk − 13)/6; the free-fall time tff ∝ 1/√ρ therefore
%scales with mass as
%tff ∝ M (13−nk )/12.
On the smallest scales n$_k~\rightarrow$ -3; %and tff ∝ M 4/3;
%a steeper relation than 
  on the largest scales %when
n$_k ~\rightarrow$ +1. %and tff ∝ M . Thereforei
Given the mass dependence of the free fall time, small clumps
have fully virialised when the violent relaxation
phase of the larger halo begins.
%Since there is no fixed scales of mass or radius
%in gravitational dynamics, all virialised structures
%obey the same relation (6). 
%If we lump together
For all those of virial radius $<$ r, of mass m %chosen
such that M/m = N $>>$ 1, we have
R/r $\propto$ ( M/m)$^\gamma$. %= N (n$_k$ +5)/6 .

%The collision rate (or periastron frequency) per unit volume, x, of masses m with central mass M
%and R = GM/c$^2$
%is x~$\propto$~n~v/r for objects in orbit with velocity v, where n is the number density of m, and so
%$$ x \propto \frac{n v}{M} \left ( \frac{M}{m} \right )^\gamma%\right$$
%~~~~~ \propto~~ \frac{ v}{m^{(3 + \gamma)} M^{(1 - \gamma)} } \eqno (2)$$%\frac({M}{m})^\gamma$$

%\noindent In the Appendix evidence is given for working with 0.3 $\lesssim~\gamma~\lesssim$ 0.5.}

For present purposes let us consider the baryons associated with two PBHs, a DM particle and an SMBH of mass 3 $\times$ 10$^7$ M$_\odot$. 
 The key to rapid star formation within this gas is its ability to cool efficiently. In the early universe, via atomic line and H$_2$ molecular cooling, the cooling time  t$_{cool}$ in these dense, virially heated gas halos can be significantly shorter than their free-fall time (t$_{ff}$). When the Rees-Ostriker criterion (t$_{cool}~ <~ t_{ff}$) is met, the gas cloud undergoes catastrophic collapse and fragments, triggering a massive starburst. The characteristic timescale for this process can be as short as a few hundred million years, essentially setting the galaxy formation clock.
Alternatively, one can consider
their Kelvin Helmholtz time
at Eddington luminosity and radius 1 AU is less than t$_2$ = 3 $\times$ 10$^8$ years independent of mass.
This is the galaxy/star formation time for galaxies formed around a 6 $\times$ 10$^7$ M$_\odot$ PBH nucleus.
Collisions are fewer by $\surd$M for more massive nuclei, and the (1+z)$^{3/2}$ factor\footnote{PBH masses follow M $\sim~ c^3 t/8G$ and the exponent comes from the scale factor a $\sim~ t^{2/3}$ in the matter dominated era.} between PBH formation
and recombination is bigger for less massive nuclei. If M8 is 10$^8$ M$_\odot$, so the scaling is t$_1$ $\propto$ (M/M8)$^{-1}$, %reaching a Gyr at M = 1.2 $\times$ 10$^8$ M$_\odot$, and 
dropping to a free-fall time of 10$^8$ years at 9 $\times$ 10$^8$ M$_\odot$.
This is a useful acceleration of massive galaxy formation time. %and a set of physical simulations
%with PBH seeds like these would be a valuable addition to current JWST-relevant galaxy formation models,
%on the unproven assumption that such massive PBH exist.
A suite of hydrodynamical simulations incorporating PBH seeds would therefore be a valuable addition to current galaxy formation models, contingent on the (as yet unproven) assumption that such massive PBHs exist. % and contribute to the dark matter budget.
\subsection{Feedback}
 Acceleration of baryons collapsing might be negated by stellar
feedback, for instance by supernova explosions of particularly short-lived stars. Again hydrodynamic simulations are the way to study this. However, one can get some insight from timescales, assembled in Table 1.

\begin{table}[H]
\caption{Free-fall and star formation timescales}
\begin{tabular}{lllrr}
	\\
\hline
Mass&T$_{ff}$&Baryons&Eddington & Efficiency\\
M M$_\odot$&Myrs&M$_\odot$&L L$_\odot$&\\
	(1)&(2)&(3)&(4)&(5)\\
\hline
100    & 100 & 18.7&       6 $\times$ 10$^5$&0.03\\
10$^4$ & 10  & 1.9 $\times$ 10$^3$& 6 $\times$ 10$^7$&0.3\\
10$^6$ & 1   & 1.9 $\times$ 10$^5$&6 $\times$ 10$^9$&3\\
\hline
\multicolumn{5}{l}{Star formation efficiency $>$ 3 Myrs /T$_{ff}$}\\
\end{tabular}
\end{table}

For three progressively larger mass PBH nuclei in column (1) a free-fall
time, $$T_{ff} = \frac {\pi R^{3/2} }{2 (M + m)^{1/2}}$$ is given
for an appropriate value of initial radius R in column (2), and the
mass m of gas accompanying M appears in column (3). If star formation
is at the Eddington rate, that is in column (5). This corresponds
to feedback limiting the star formation. But there is a period
before the first supernova occurs of order 3 Myrs for the most massive stars.
The ratio of this time to T$_{ff}$ is effectively the star formation
efficiency in the presence of feedback. This is sufficently high
that large PBH nuclei are able to trigger high luminosity star formation
before being quenched by feedback. Hydrodynamic simulations are needed
to test this presumption, and those that have been run up to now
suggest that feedback does not negate star formation (Liu \etal 2025),
or merely hampers or regulates it (Zhang \etal 2025).

 The demands of very high star formation efficiency, possibly even in excess of the cosmic baryon mass budget in collapsed structures, are discussed by 
 Liu \& Bromm (2022).
%\section{The galaxy lumino
\subsection{Alternatives}
%{\color{red}
PBH acceleration of galaxy formation is just one of a number
of mechanisms that have been presented to respond to the challenge of
early massive galaxies, and, beyond that, early passive galaxies (Glazebrook
\etal 2024; Koulen, Profumo \& Smyth 2025). Direct collapse black holes (Vikaeus, Whalen \& Zackrisson 2022)
do all that PBH halos can do
to provide a deep potential to host star formation, and also to seed AGN.
Runaway baryonic collapse, which has long been a favored model for globular
cluster formation, may occur independently of DM interaction. And a whole
chapter is under development (Ellis 2025) for the role of Population III's massive stars
as a scenario for JWST's higher redshift galaxy discoveries.%}
\subsection{Diffuse Galaxies}
If PBH nuclei play a pivotal role in galaxy formation, there is a corollary.
Protogalaxies $without$ a PBH nucleus should be different. %fo
%The default model for galaxy formation assumes a uniform mix of baryon and dark matter as input. However, it is easy to imagine exceptions. For example, the DMG toy model for PBH formation predicts up to a 64\% conversion of energy into PBH of various sizes in one generation. But a second generation would raise that percentage to 87\% ie f(2-f) for f=0.64. These different percentages characterise regions smaller but not much smaller than the horizon, leading to shallow potentials and larger galaxy radii in the baryon deprived regions. A smaller scale unrelated cause of such deprivation would be early cluster formation  
%depleting the IGM around still  forming galaxies.
%Confusingly, the other way for a galaxy to lose mass during its collapse is to have very little dark matter of the mass conserving kind,
%WIMPs, Planck mass relics and planet mass PBH.
%Main sequence stars have dlogm/dlogt $\approx$ -1, more or less independent of mass. If they commence fusion early in the galaxy's collapse,
%this may also lead to thinly spread baryons in a shallow potential
%of abnormally large radius.
%\section{Do PBH also play a role in determining galaxy masses?}
%Detritus theory of UDGs
%Close cousin of the failed galaxy theory
It's an interesting coincidence that at the free fall time in the universe, 10$^8$ years or z = 25, the pregalactic uniform density of gas is
$\rho$ = 1.88 $\times$ 10$^{-29} ~\Omega_b h^2 (1+z)^3$ gm/cc.
%= 1.88/2 3.08^3 10^-7 0.0224 26^3 Msun/pc^3
%=1.88 3 10^-6 0.0224 17576 Lsun /pc^3 for M/L=1 (old stellar pop)
For an old stellar population,
taking a 1 kpc thick slice perpendicular to the observer, the
surface brightness is 23.7 mag/arcsec$^2$ in V.
This suggests Ultra Diffuse Galaxies, lacking a PBH nucleus, (UDGs, Forbes \etal 2020) may be the leftovers from galaxy formation. 

If regular galaxy formation is seeded by $\sim$10$^6$ M$_\odot$  supermassive PBHs, which create the gravitational field (density inhomogeneity) for collapse, those regions without one will be left over as UDGs. 
%Why do UDGs form stars?
Local inhomogeneities collapse after 10$^8$ years. If the filling factor of star formation is low, the surface brightness will be that much fainter. 
The Analysis of Galaxies at the Extremes project
(Forbes \etal 2025) has asked, How and when can UDGs have globular clusters?
Because they are so close to pregalactic background density, UDGs have very shallow potential wells. If globular clusters are pregalactic (Mould \& Hurley 2025), from say z= 100, they wander around freely. The somewhat deeper UDGs will gain globular clusters. The shallowest ones will lose them. 
``Wandering" black holes are recognized entities (Sturm \etal 2026).
Another interesting test is whether UDGs harbor  detectable SMBHs.
%The thesis in this section would say, no.
X-ray and radio surveys can give a statistical result (see Mirakor \& Walker 2021).
Finally, nothing here precludes HI rich or even dark galaxies from being classified as UDGs and supporting this theory 
(Benitez-Llambay \etal 2024;  O'Beirne \etal 2025).
%What sets the size of UDGs?
%Tidal interactions with regular galaxies. 
%Can UDGs survive in x-ray clusters?
%If they have completed star formation, they are not disturbed by the hot IGM. 
Are UDGs uncollapsed galaxies, lacking a PBH nucleus ? This requires further simulations without PBH nuclei to analyze the halos that develop in this form.
\section{Conclusions}
\begin{itemize}
	\item The presence of PBH seeds of order 10$^8$ M$_\odot$ can shorten
		the galaxy formation time to 10$^8$ years, z~$\approx$~30, providing manoeuvering room for models to fit JWST massive young galaxy discoveries.
	\item  PBHs are not unique as accelerators of galaxy formation.
Gravitational potentials may be deepened by direct collapse of matter into black holes or by baryonic processes particular to Population III. These 
		mechanisms are also candidates.
	\item  With time dependent dark energy now contemplated,
		it is also possible that older ages for the universe
		can accommodate JWST's stellar ages
(Batic, Medina \& Nowakowski 2024).
\item Lower mass PBHs may be candidates for a significant fraction of halo DM. These evolve to resemble the halos of normal galaxies.
%	\item The expectation from mergers of these halos is a Schechter mass or luminosity function.
	\item The residue from these processes without a massive PBH nuclear seed can be compared with the new class of UDGs.
\end{itemize}
%\begin{figure}[H]
%\includegraphics[width=.55\textwidth]{newell.pdf}
%	\caption{Radial profiles  obtained by azimuthally averaging
%	the mass density  of the three halos described in the text (left, with radii in pixels) and their potentials in log velocity squared dimensions (right, with radius in arbitrary units).  This shows the depth
%	of the potential in run 135 relative to runs 137 and 138. Salucci (2019) has noted the extensive DM cores in diffuse galaxies,  and this is evident in the central flatness of these profiles. }
%\end{figure}

%\begin{figure}[H]
%\includegraphics[width=.55\textwidth]{imf.pdf}
%\caption{PBH mass functions for two of the galaxies in the previous figure.}
%\end{figure}
%\section{A short history of AASTeX} 

\section*{References}
%\reftitle{References}
\noindent %Aaronson, M. Mould, J.\& Huchra, J. 1979, ApJ, 229, 1\\
Ade, P. \etal , Planck collaboration 2020, A\&A, 641, A6\\
Batic, D., Medina, S. \& Nowakowski, M. 2024, IJMPD, 34, 255044\\
Batten, A. \& Mould, J. 2025, submitted to MNRAS\\
Benitez-Llambay, A. \etal 2024, ApJ, 973, 61\\
Bicknell, G. \& Henriksen, R. 1979, ApJ, 232, 670\\
Binney, J. \& Knebe, A. 2002, MNRAS, 333, 378\\
Boylan-Kolchin, M. 2025, MNRAS, 538, 3210\\
Buzzoni, A. 2002, AJ, 123, 1188\\
Buzzoni, A. \etal 2009, {\it New Quests in Astrophysics, II UV properties of evolved stellar populations,} Springer, eds. M. Chavez, E. Bertone, D. Rosa-Gonzales \&
L. Rodriguez-Merino\\
%Alcock, C. \etal 1993, Nature, 365, 621\\
%Alonso-Monsalve, E. \& Kaiser, D. 2023, Phys Rev L, 132.231402\\
Capelluti, N. 2023, APS April Meeting, abstract id.B02.002\\
 Carr, B., Kohri, K., Sendouda, Y. \& Yokoyama, J. 2016, PRD 94.044029\\
%Carr, B. \& K\"uhnel, F. 2021, arxiv 21100282\\
Carr, B. \& K\"uhnel, F. 2021, SciPost Phys. Lect. Notes 48, arxiv 211002821\\
Carr, B. \etal 2021, Rept. Prog. Phys., 84(11), 116902\\
%Courtois, H., Mould, J., Hollinger, A., Dupuy, A. \& Zhang, C. 2025, A\&A, 701, 187\\
Cuillandre, J-C. \etal 2024, A\&A, 697, 11\\%, arxiv 2405.13501\\
%Dav\'e, R., Thompson, R., \& Hopkins, P. 2016, MNRAS, 462, 3265\\
Dayal, P. \& Maiolino, R. 2025, arXiv 2506.08116\\
%da Cunha, E. \etal 2012, IAUS, 284, 292\\  
Delos, M. S. \etal 2024, JCAP, 12, 5\\%arxiv 2410.01876\\
de Graaff, A. \etal 2025, Nat As, 9, 280\\ 
Ellis, R. 2025, arXiv 2508.16948\\
Forbes, D. \etal 2020, MNRAS, 494, 5293\\
Forbes, D. \etal 2025, MNRAS, 536, 1219\\
Gannon, J. \etal 2022, MNRAS, 510, 946\\
Glazebrook, K., Nanayakkara, T., Marchesini, D., Kacprzak, G. \& Jacobs, C. 2024, IAUS, 377, 3\\
Green, A. 2024, Nuclear Physics B, 1003, id.116494\\
Hawking, S., 1971,  MNRAS, 152, 75\\
Hoshi, A. \& Yamada, T. 2025 ApJ, 988, 234\\
Kaiser, N. 1986, MNRAS 222, 323\\
%Key, R. \etal 2024, submitted to ApJ\\
Kohri, K. 2024, "Overview on Current Constraints on the Primordial Black Hole Abundance",
{\it Primordial Black Holes}, ed Christian Byrnes, Springer, p.497\\
Koulen, J., Profumo, S. \&  Smyth, N. 2025, arxiv 2506.06171\\
%Kourkchi, E. \etal 2022, MNRAS, 511, 6160\\
K\"uhnel, F. 2025, "Positive Indications for Primordial Black Holes",
{\it Primordial Black Holes}, ed Christian Byrnes, Springer, p.453\\
Inman, D. \& Ali-Ha\"imoud, Y. 2017, arxiv 1907.08129\\
Liu, B. and Bromm, V. 2022 ApJL 937, L30\\
Ma, Y.,  Greene, J. \& Setton, D. 2025, arxiv 250408032\\
%McGaugh, S. \etal 2000, ApJ, 533, L9\\%\& Bothun, G.\\a
Merlin, E. \etal 2024, EAS2024, Annual Meeting, July, Padova, Italy\footnote{https://eas.unige.ch/EAS2024/}\\%. Session S3 : New light on Galaxies from Cosmic Dawn to Noon, Contributed talk, id. 567
Mirakor, M. \& Walker, S. 2021, MNRAS, 503, 679\\
Mould, J. 2025a, RNAAS, 9, 103\\%in press\\
Mould, J.  2025b, ApJ, 984, 59\\% arxiv 2504.11595\\
%Mould, J. \etal 2024, MNRAS, 533, 925\\
Mould, J. \& Batten, A. 2025, arxiv 2507.11023\\
Mould, J. \& Hurley, J. 2025, arxiv 2509.02165\\
Niikura, H. \etal 2019, Nature Astronomy, 3, 524\\
O'Beirne, T. \etal 2025, MNRAS, 544, 1799O\\ 
Padmanabhan, T. 1993, Structure formation in
the Universe, Cambridge University Press \\
P\'erez-Gonz\'alez, P. \"Ostlin, G. \& Costantin, L. 2025, ApJ, 991, 179\\%arxiv 2503.15594\\
Planck collaboration 2020, A\&A, 641, A6\\
%Moniez, M. 2001, Proc. XXXVth Rencontres de Moriond, Tran, Mellier \& Moniez. Les Ulis: EDP Sciences\\
%Paczynski, B. \etal 1996, IAU Symposium, 169, 93\\
%Press, W. \& Schechter, P. 1974, 187, 525\\
 %   The SEEDZ Simulations: Methodology and First Results on Massive Black Hole Seeding and Early Galaxy Growth
Prole, J. \etal 2025a, %ohn A. Regan, Daxal Mehta, Rudiger Pakmor, Sophie Koudmani, Martin A. Bourne, Simon C.O. Glover, John H. Wise, Ralf S. Klessen, Michael Tremmel, Debora Sijacki, Ricarda S. Beckmann, Martin G. Haehnelt, John Brennan, Pelle van de Bor, Paul C. Clark
 arXiv 2511.09640\\ 
Prole, L. \etal 2025b, OJAp 8, 126\\%arxiv 2506.11233\\
%Salucci, P. 2019, Astronomy \& Astrophysics Review,  27,  id. 2\\%, 60 pp.
%Schechter, P. 1987, ApJ, 203, 297\\
Sobrinho, J. \& Augusto, P. 2024, MNRAS, 531, L40\\
Somerville, R. \& Primack, J. 1999, MNRAS, 310, 1087\\
Sturm , M. \etal 2026, ApJ, 996, 4\\%1rxiv 2511.09461\\
Tran, T., Geller, S., Lehmann, B. \& Kaiser, D. 2024, Phys Rev D, 110.063533\\ 
Sugiyama, S., Takada, M., Yasuda, N. \& Tominaga, N. 2026, arxiv 2602.05840\\
%Tully, R.B. \& Fisher, J. 1977, A\&A, 54, 661\\%Taylor, Q. \etal 2024, PhysRevD, 109, 406\\ 
%Wang, Q. \etal 2021, Phys Rev D, 104, 8, id.083546\\
Vikaeus, A., Whalen, D. \& Zackrisson, E. 2022, ApJ, 933, L8\\
Xiao, M. \etal 2024, Nature, 635, 311\\
Yung, L., Somerville, R. \& Iyer, K. 2025, MNRAS, 543, 3802\\% arxiv 2504.18618\\
Zel'dovich, Y.B. \& Novikov, I.D. 1967, Sov. Astron. 10, 602\\
%Zhang, S. \etal 2025, arxiv 2506.11233\\
Zhang, S., Liu, B., Bromm, V. \& K\"uhnel, F. 2025, arxiv 2512.14066\\
Zhang, S., Liu, B., \& Bromm, V. 2025, arxiv 2512.11381\\
%Zheng, R. \etal 2022, Chinese Physics C, 46, 4, id.045103\\
Ziegler, J.,  Freese, K., Lozano, J., \& Montefalcone, G. 2025, arxiv 2507.21409\\

%\acknowledgments{
%The ARC Centre of Excellence for Dark Matter Particle Physics is funded by the Australian Research Council. Grant CE200100008. 

\noindent 
%\begin{tabular}{@{}ll}
%MDPI & Multidisciplinary Digital Publishing Institute\\
%DOAJ & Directory of open access journals\\
%TLA & Three letter acronym\\
%LD & Linear dichroism
%\end{tabular}
%}

\appendix
\renewcommand{\thetable}{A\arabic{table}}
\renewcommand{\thefigure}{A\arabic{figure}}
\setcounter{figure}{0}
\setcounter{table}{0}
%\vpar
%\vspace{-1.5in}
%\vpar
%\section*{Appendix}
\subsection*{Parameter summary}
	The dynamics and density of the all-PBH DM-only simulations depend on a small number of parameters
	which are varied in Table A1 to show these dependencies. Here $\sigma$ is the 3D velocity dispersion of the n particles, and r$_e$
	is the effective radius, the median distance of the particles from the nucleus. The simulations are scale free
	which means that ratios of the quantities are meaningful, but the masses, for instance, are not specifically solar masses.
\begin{table}[H]
%\vspace{-2cm}
\caption{Relationship to parameters}
	\begin{adjustwidth}{-1in}{}
\begin{tabular}{|lrrrr|rrrrlll|}
\hline
Parameter& \# & Particles&$\sigma$ & r$_e$ & \# &Particles& $\sigma$ & r$_e$ & Relation&m1, m2 & Notes \\
Particles n& 135& 10000 & 207 & 89 & 136 & 50000& 1035 & 587 & $\sigma~\propto$ n&0.1, 1 & 1\% mass loss\\
m1, m2 & 138& 50000& 1.8* &586 & 139 &250000& 1.9*&334&$\sigma~\propto~\surd$ m1 m2&10$^{-6}$, 10$^{-5}$ & 1\% \\
Mass loss  &142& 200000 & 1195  &240    & 144 &200000 &638 &126&$\sigma~\sim~$1/n\% &0.01, 0.1& 2\%, 0\%; 1000$^\dagger$\\
Nucleus&236&50000&345&137&237&50000&345&137&$\sigma$ independent&1000: 1&\\
\hline
\multicolumn{12}{l}{Notes: Runs 236 \& 237 were run with different gravity softening.}\\ 
\multicolumn{12}{l}{*$\times$ 10$^{-6}$~~~~~~~~~~$^\dagger$nucleus}\\
\end{tabular}
	\end{adjustwidth}
%    \vspace{-2cm}
	\end{table}

%    \columnbreak
%    \vfill\eject
%    \clearpage
%\begin{multicols}{2}%    
%    \twocolumn
	The full simulation specification is an unlimited box size\footnote{Reflective boundary conditions have been tested, but have a marginal effect
	for the parameters listed above.}, and total particle number, particle mass ratio(s) as above. Gravitational softening is obtained by adopting
	a minimum particle separation of 10$^{-10}$ units;  starting redshift is z = 100; Planck (2020) cosmological parameters were adopted.
	The PBHs are inserted in a  uniform density sphere at rest.%(IC placement and velocity).
%\pagebreak
	
    Figure A1 shows the effect of  massive PBHs on the radial profile.
    The mass these add to the nuclear region forms a cusp. Run 61 has low mass DM and forms a core. Figure A2 shows the radial profiles of the runs in Table A1. The effect of nuclear mass on radial profile is seen in detail in Figure A3. The transition from core to cusp with increasing density near the nucleus, takes place at 3 $\times$ 10$^5$ M$_\odot$.
      %\documentclass{article}
%\usepackage{tikz}
%\begin{document}
\begin{figure}
%\vspace{-3in}
\begin{tikzpicture}

    \begin{scope}
    \node {A2: \includegraphics[width=5 in]{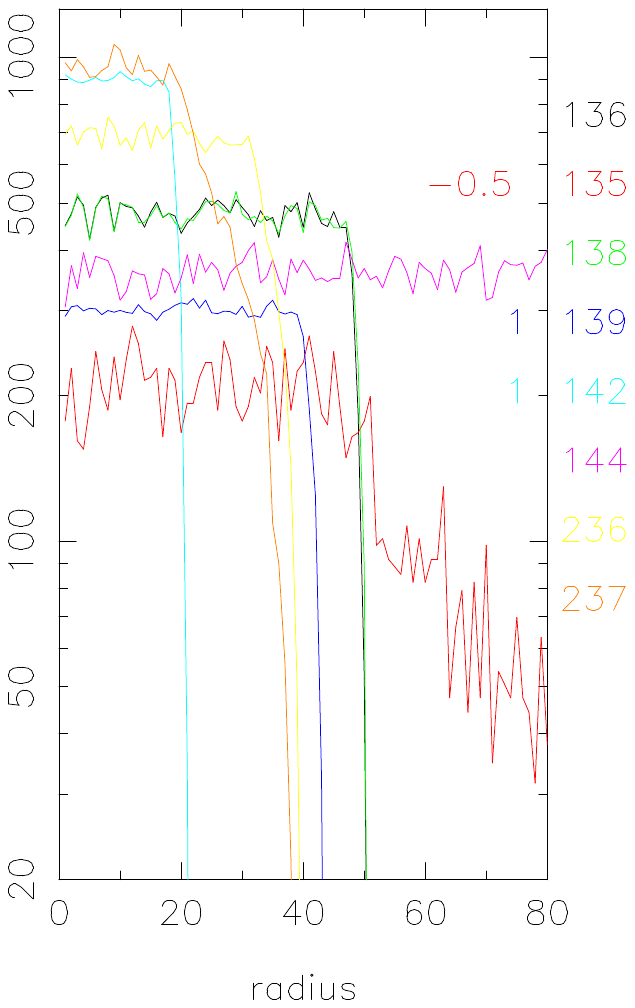}};
    \end{scope}
    \begin{scope}[xshift=-6cm]
    \node {A1: \includegraphics[width=5 in]{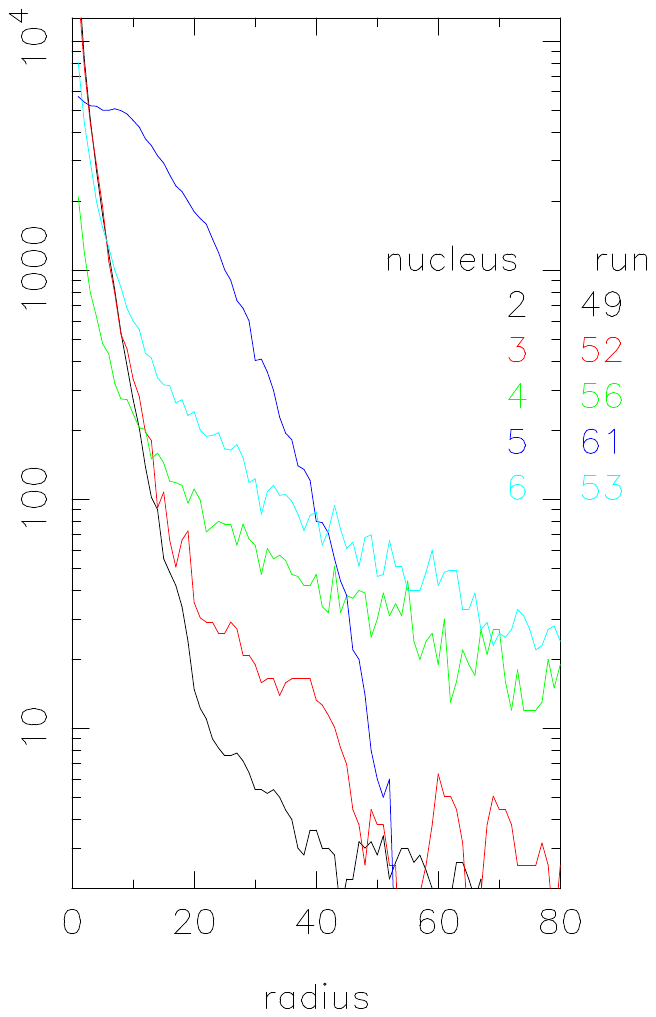}};
    \end{scope}
        \begin{scope}[xshift=6cm]
    \node {~~~A3:\includegraphics[width=5 in]{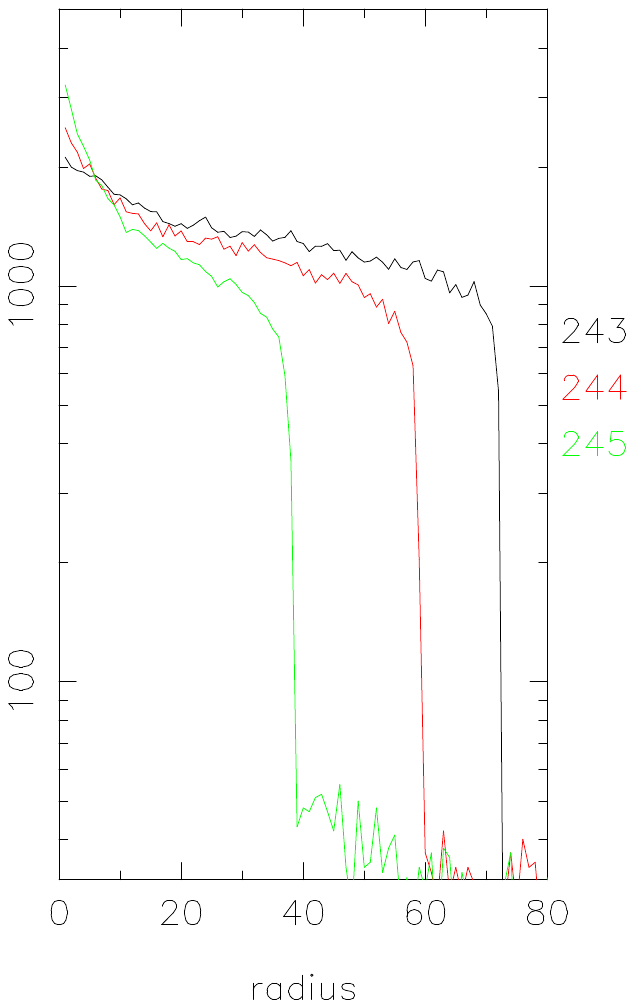}};
    \end{scope}
\end{tikzpicture}
%\vspace{-2in}
%\begin{figure}
\caption{The effect of  mass around the nucleus on radial profile. These are all cusps with increasing density near the nucleus, with the exception of run 61, which has m$_1$ and m$_2$ five orders of magnitude below the nuclear mass.}

\caption{For clarity runs 139 and 142 have been shifted up by a factor of 10
and run 135 have been shifted down by a factor of $\surd$10}%
\caption{Close up of the effect of nuclear mass on the cusp/core morphology. A 10$^5$ M$_\odot$ nucleus
(black) has a core; a 300,000 M$_\odot$ nucleus
(red) is growing a cusp; a 10$^6$ nucleus has a strong cusp.}
%\end{minipage}
%\end{tikzpicture}
%\vspace{-2in}
\end{figure}
%\end{document}

	%\end{figure}
    %\documentclass{article}
%\begin{document}
%\renewcommand{\thetable}{A\arabic{table}}
%\renewcommand{\thefigure}{A\arabic{figure}}
%\setcounter{figure}{0}
%\setcounter{table}{0}
%    \centering\
%\captionsetup[table]{
 %   justification=raggedleft, % Align caption text to the right
  %  singlelinecheck=false      % Prevent centering for short captions
%}
%\caption{%\Huge \bf 

\begin{table}%\Huge
%\raggedleft
\centering

\begin{tabular}{lrrr}
\multicolumn{4}{l}{{\bf Table A2:} Details: %of runs in 
Figs A1 \& A3}\\
\hline
run&nucleus&~~~m1~~~&~~~m2\\
\# &M$_\odot$&M$_\odot$&M$_\odot$\\
49&100&1~~~&100\\
52&1000&1~~~&1000\\
53&10$^6$&1000&10$^4$\\
56&10$^4$&1000&10$^4$\\
61&10$^5$&10$^{-5}$&1\\
132&10$^7$&10&100\\
185&6 $\times$ 10$^6$&50&600\\%small dir 135
187&6 $\times$ 10$^6$&5&50\\%small dir 137
188&6 $\times $10$^4$&50&600\\%small dir 138
243&%1 $\times$ 
10$^5$&10$^{-4}$&10\\%$^{-4}\\
244& 3 $\times$ 10$^5$&3 $\times$ 10$^{-4}$&30\\
245& 10$^6$&10$^{-4}$&100\\
\hline
\end{tabular}
\end{table}
%\end{document}

\subsection*{PBH constraints}
Although there are constraints on massive PBHs, shown in Figure A1, these
do not impact the present model. They do not rule out PBHs of mass
10$^5$ to 10$^7$ M$_\odot$ (37.3 to 40.3 in log M [g]). They simply
tell us that only a small amount of the total DM is at these masses.
Not shown here is the ``asteroid window" (10$^{17}$ to 10$^{25}$ grams),
where the bulk of the DM may conceivably lie in PBHs. 
\begin{figure}%[H]
%\vspace{-3.5in}
\includegraphics[width=.7\textwidth]{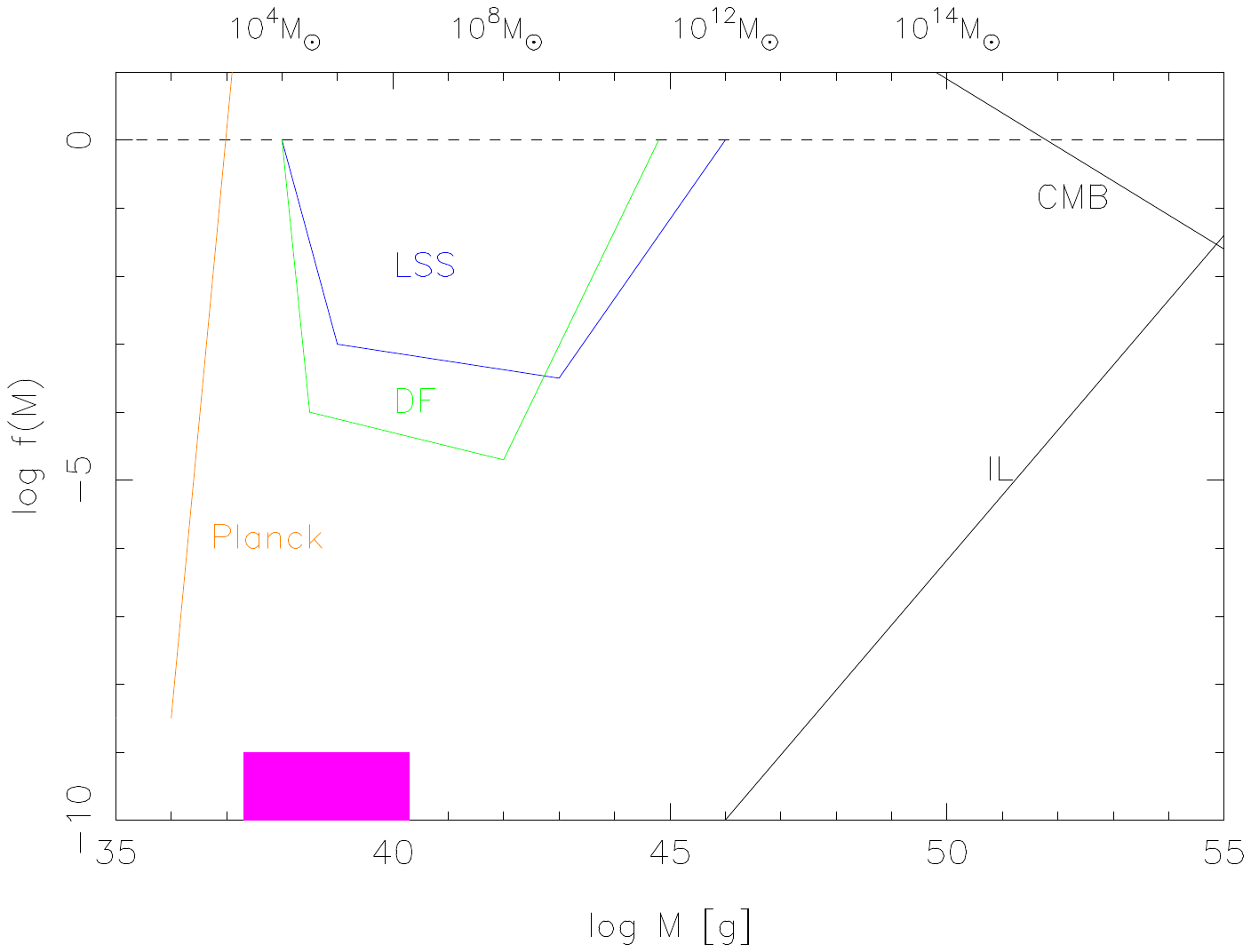}
	\caption{Limits on the fraction f(M) of the DM that can be PBHs of mass M. The figure is reproduced in the relevant part from Kohri (2024), who updated it from
	Carr \etal (2021). Areas that are ruled out are to the right of the 
	labels CMB and IL. Other areas ruled out in blue and green
	are due to dynamical friction and large scale structure constraints,
 and Planck anisotropy limits are shown in orange.
	References are given by Kohri (2024). The pink rectangle
	shows the range 10$^4$ -- 10$^7$ M$_\odot$ which is of interest in the present work. %, located
%	so that they would satisfy this constraint. 
	The `incredulity limit'
	(IL) corresponds to one hole per Hubble volume.}
\end{figure}

%% Please use the acknowledgment and contribution environments. This will 
%% be anonomyized when the "anonymous" style option is used. 
\begin{acknowledgments}
%section*{Acknowledgments}
Simulations were carried out on 
Swinburne University's Ozstar \& Ngarrgu Tindebeek supercomputers, the latter named by Wurundjeri elders and translating as "Knowledge of the Void" in the local Woiwurrung language. 
Thanks to Jonah Gannon for helpful discussions %.%\subsection*{Code availability}
on UDGs. Thanks to the referees who advised on improvements to the paper.%}
%\conflictsofinterest{The author declares no conflicts of interest.
%}
%The evolutionary tracks and toy model with README files
 %are available at https://github.com/jrmould

%\funding{
The ARC Centre of Excellence for Dark Matter Particle Physics is funded by the Australian Research Council. Grant CE200100008. %Simulations were carried out on 
%}

\end{acknowledgments}

\end{document}